# Magneto-optical characterization of GeSn and GeSn/SiGeSn heterostructures

Mehdi Ashrafganjoie[1], Kelvin Dsouza[2], Prithu Bhatnagar[2], Khalid Hossain[3], Mangal Dhoubhadel[3], Preston Webster[4], Simone Assali[5], Patrick Daoust[5], Oussama Moutanabbir[5], Shui-Qing Yu[6], Daryoosh Vashaee[2,7], Demitry Farfurnik[1,2*]

[1] Department of Physics and Astronomy, North Carolina State University, Raleigh, North Carolina 27607, USA
[2] Department of Electrical and Computer Engineering, North Carolina State University, Raleigh, North Carolina 27606, USA
[3] JP Analytical, LLC., Richardson, Texas 75081, USA
[4] Air Force Research Laboratory, Space Warfare Directorate, Kirtland AFB, New Mexico 87117, USA
[5] Département de Génie Physique, École Polytechnique de Montréal, Montréal, C.P. 6079, Succ. Centre-Ville, Montréal, Québec, Canada H3C 3A7
[6] Department of Electrical Engineering, University of Arkansas, Fayetteville, Arkansas 72701, USA
[7] Department of Materials Science and Engineering, North Carolina State University, Raleigh, North Carolina 27606, USA

Hole spin qubits in germanium (Ge)-based heterostructures have demonstrated their potential for scalable quantum information processing using all-electrical gate operations. Furthermore, the emerging material platform of germanium-tin (GeSn) can feature a direct bandgap, which makes it promising for establishing spin-photon interfaces for quantum networking. Here, we perform magneto-photoluminescence measurements of a $Ge_{0.88}Sn_{0.12}/Si_{0.02}Ge_{0.89}Sn_{0.09}$ double quantum well using the double modulation Fourier transform infrared-based photoluminescence spectroscopy. Our measurements reveal theoretically expected diamagnetic shift at low magnetic fields as well as the linear trend of zeroth-level Landau quantization at higher fields and Zeeman-induced polarization-dependent energy shifts at ± 12 T. We extract an effective g-factor of ~ 2 and an excitonic reduced mass of ~ 0.04 $m_e$ consistent with previous estimations for heavy-hole Γ-valley excitons. The observation of sizable Zeeman splitting is consistent with strong spin-orbit interaction in Ge-based hole systems, which can enable electrically driven spin control. Our analysis can be adopted for studying and evaluating group-IV semiconductor heterostructures as hosts for hole spin qubits toward scalable quantum information processing.

## I. INTRODUCTION

Semiconductor spin qubits in gate-defined quantum dots enable scalable quantum information processing via fast, all-electrical coherent control [1]. Specifically, holes in Germanium (Ge) quantum dots have emerged as a promising computational platform [2-4] due to the p-orbital character of the Ge valence band that strongly reduces the coupling to nuclear spins [5], leads to reduced decoherence for hole spins relative to electron spins, making isotopically purified Ge [6, 7] an attractive platform for high-coherence spin qubits [8, 9]. However, a fundamental limitation of Ge is its indirect bandgap, which severely restricts the potential use of this material as an interface between single spins and single photons for quantum networking [10, 11].

To overcome the limitations imposed by the indirect band gap of Ge, germanium-tin ($Ge_{1-X}Sn_X$) alloys have attracted considerable attention as material platforms for hosting spin qubits. These alloys undergo an indirect-to-direct band gap transition for tin (Sn) concentrations X ≥ 6.5% as predicted theoretically for a fully relaxed material [12-15] and confirmed experimentally [16]. Over the past several decades, $Ge_{1-X}Sn_X$ has been extensively studied for its tunable band gap, high carrier mobility, and compatibility with group-IV semiconductor technology, making it a promising platform for both electronic and optoelectronic applications [14]. $Ge_{1-X}Sn_X$-based heterostructures exhibit strong, electrically tunable spin-orbit coupling with a significant Rashba contribution arising from gate-induced structural inversion asymmetry, where symmetry considerations lead to a predominantly cubic momentum dependence in contrast to many III-V systems [17]. Magneto-transport measurements revealing weak anti-localization and large spin splitting confirm enhanced cubic Rashba coefficients in $Ge_{1-X}Sn_X$ compared to pure Ge, enabling efficient all-electrical spin control via electric-dipole spin resonance [18]. The efficiency of the control is expected to be further strengthened in quasi-one-dimensional geometries such as nanowires, which makes them promising as scalable quantum processors and spin-photon interfaces [19].

* Contact author: dfarfur@ncsu.edu

In addition to the potential for spin-photon interfaces applications, the emergence of a direct bandgap in $Ge_{1-X}Sn_X$ provides a unique opportunity to probe its optoelectronic and magneto-optical properties using well-established, non-destructive techniques such as photoluminescence (PL) spectroscopy [20]. Magneto-optical measurements, in particular, enable direct investigation of spin-related properties relevant to quantum computation, including effective g-factors and spin splitting under applied magnetic fields [21-22]. Although temperature-dependent PL has been extensively used to confirm the direct band-gap nature of $Ge_{1-X}Sn_X$ alloys [23-25], systematic magneto-PL investigations of high-Sn-content of $Ge_{1-X}Sn_X$ heterostructures, especially under high magnetic fields, remain largely unexplored [26, 27].

Despite the direct bandgap of $Ge_{1-X}Sn_X$, the PL signals emitted from such alloys in the mid-wave infrared (IR) region of the electromagnetic spectrum are weak due to non-radiative decay pathways. To measure comparably weak PL signals emitted from other materials, researchers have utilized Fourier Transform Infrared (FTIR)-based PL spectroscopy while the FTIR spectrometer operates in a “step-scan” mode [28-31] integrated with a lock-in amplifier to implement the phase-sensitive detection scheme for enhancing the signal to noise ratio by suppressing the thermal radiation background. In the absence of step-scan mode, the alternative approach to implement phase-sensitive detection for performing PL measurements using the commonly available “rapid scan” mode of the FTIR spectrometer is the double modulation (DM) approach [32, 33]. In this approach, the scanning mirror speed is sufficiently reduced to produce an interferogram whose frequency content is significantly slower than the pump laser’s (and thus the PL emission’s) repetition rate, enabling the lock-in amplifier to reproduce the slowly varying interferogram at its output.

Here, we construct an experimental setup to implement a DM FTIR-based PL spectroscopy for the magneto-optical characterization of $Ge_{1-X}Sn_X$ heterostructures. Our approach involves the optimization of parameters such as the frequency of double modulation, the velocity of the scanning mirror, the laser power, and the suitable electro-optical modulator for high-power laser beams. After successfully characterizing previously studied bulk samples using our analysis approach, we apply it to study the temperature-dependent PL and the first magneto-PL from a double-quantum-well (DQW) sample with high Sn concentration. The temperature-dependent PL measurements reveal thermal phonon-independent spectral broadening and magneto-PL measurements under external magnetic fields of up to 12 T exhibit theoretically expected diamagnetic shift, Landau shift, and Zeeman splitting. We extract the characteristic parameters of excitons in the heterostructure such as the binding energy, reduced mass, and an effective g-factor of ~ 2, thereby making our study useful for the analysis of heterostructures promising for hosting hole spin qubits.

## II. METHODS

We perform FTIR-based PL spectroscopy [34] using the experimental system illustrated in FIG. *1*. In this method, we mount samples inside a 2-Kelvin cryostat with an external magnetic field in the Faraday geometry of up to 12 T (Quantum Design, Physical Properties Measurement System, DynaCool 12). To excite the samples, an electro-optical modulator with high damage threshold (ConOptics 350-160) driven by an arbitrary waveform generator (Agilent 33220A) modulates beams of a continuous wave 980 nm diode laser (SkyEra TY-981+/-01NM-60.0W-25C-105-0.22NA) at the frequency of 38 kHz with a square wave shape and 50% duty cycle. An 1180 long-pass dichroic mirror (Thorlabs DMLP1180R) mounted on a cage system reflects the laser beam, which is then focused on the sample using a 20 mm focal length $CaF_2$ plano-convex lens (Thorlabs LA5315, NA ≈ 0.32). The emission from the sample is collected with the same lens and passes through the long pass dichroic mirror. The emission from the sample is detected and analyzed by an FTIR spectrometer in a rapid scan mode (Mattson Instrument Galaxy 5020) equipped with a nitrogen-cooled InSb detector (Infrared Associates IS-2.0) and the signal is processed through a lock-in amplifier (Zurich Instrument HF2LI) in order to suppress thermal background emission [32, 35].

## III. RESULTS

### A. Temperature-dependent PL study of bulk $Ge_{0.852}Sn_{0.148}$ sample

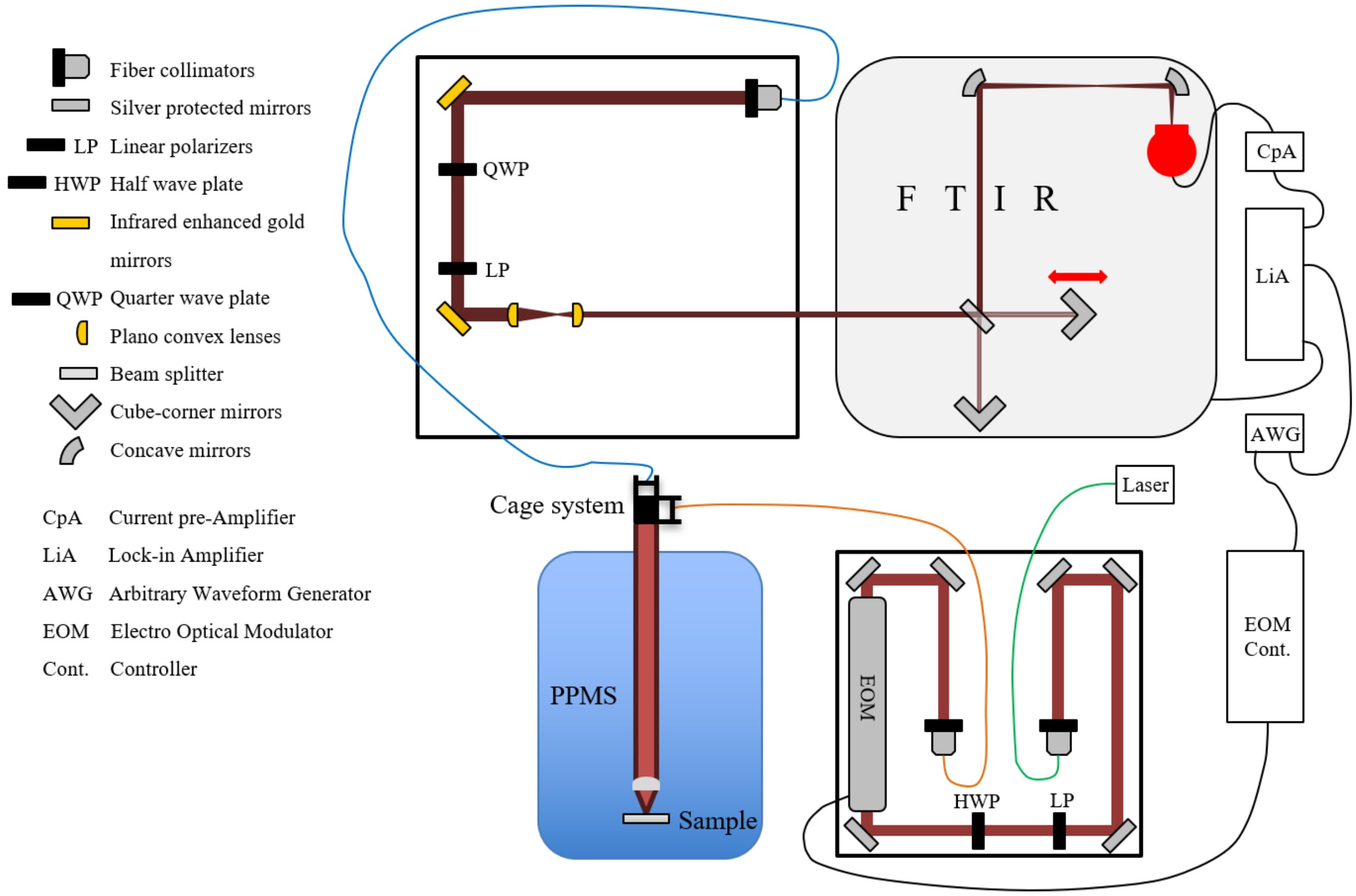

FIG. 1. The schematic of our magneto-optical characterization setup based on a DM of an FTIR spectrometer in a rapid scan mode; FTIR: Fourier Transform Infra-Red spectrometer, PPMS: Physical Properties Measurement System.

We first characterize a reference sample by measuring temperature-dependent PL from a bulk $Ge_{0.852}Sn_{0.148}$ [29, 36, 37] and compare it to measurements of similar samples characterized by other researchers that utilized an FTIR in a step-scan mode [30].

In the extracted spectra, the PL intensity sheds light on the dominant recombination mechanisms in the material and the bandgap, while the full-width-at-half-maximum (FWHM) reflects the spectral broadening of the emission and is influenced by carrier occupation within the bands, as well as by electron-phonon interactions, impurities, lattice defects, and dislocations. FIG. 2a shows the temperature-dependent spectra from the $Ge_{0.852}Sn_{0.148}$ sample at various temperatures between 40 K and 90 K, which are fitted with a single Lorentzian line shape [38]. These normalized spectra illustrate the temperature-dependent evolution of the PL emission, which exhibit an improved signal-to-noise ratio (SNR) and a higher quality fitting at lower temperatures.

The integrated PL intensity, the photon energy at the maximum intensity, and FWHM as a function of temperature are plotted in FIG. 2b, FIG. 2c, and its inset, respectively. The decrease in PL intensity with increasing temperature (FIG. 2b) indicates an increasing contribution from non-radiative recombination processes, consistent with the activation of thermally induced quenching pathways [39]. In the temperature range of this study, the FWHM remains approximately constant up to ~ 80 K (inset of FIG. 2c), suggesting that spectral broadening is independent of thermal phonons in this range. The apparent deviation at 90 K is not considered reliable due to the low SNR of the corresponding spectrum (topmost spectrum in Figure 2a), which limits the accuracy of the fitting procedure. As shown in FIG. 2c, the photon energy at maximum intensity increases with decreasing temperatures and reaches saturation at temperatures lower than 60 K, consistent with the temperature dependence of the bandgap energy. The observed temperature dependent trends of the integrated PL intensity, the photon energy, and the FWHM are in good quantitative agreement with previous measurements of PL from $Ge_{1-X}Sn_X$ alloys [30]. This quantitative agreement confirms the validity of our experimental and analysis method, enabling us to study temperature-dependent PL and magneto-PL from novel $Ge_{1-X}Sn_X$ heterostructures.

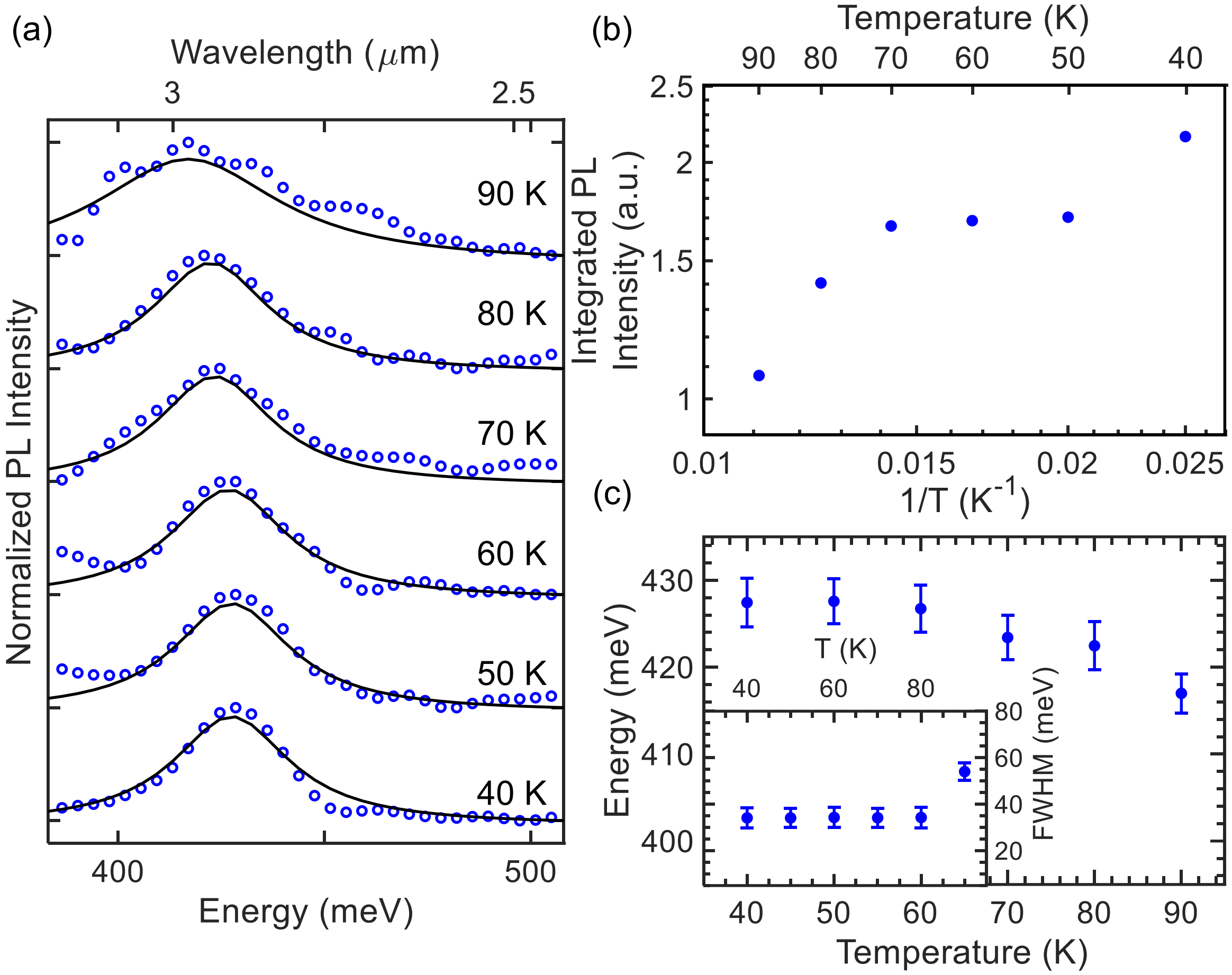

FIG. 2. The temperature-dependent PL spectra from bulk $Ge_{0.852}Sn_{0.148}$ using DM rapid-scan FTIR-based spectroscopy. (a) The normalized spectra at the temperature range 40-90 K (circles) and the fitted Lorentzian line shapes (solid lines). (b) The integrated PL intensity as a function of temperature. The decreasing intensity with temperature is consistent with thermally activated nonradiative quenching commonly observed in direct-gap GeSn. (c) The energy of the peak PL and FWHM (inset) extracted from the Lorentzian fits in (a) as a function of temperature exhibits a red shift expected in direct bandgap semiconductors and the constant FWHM (except for 90 K due to poor SNR) indicates thermal phonon-independent broadening. The error-bars are 95% confident interval of the fit procedure.

### B. PL studies on $Ge_{0.88}Sn_{0.12}/Si_{0.02}Ge_{0.89}Sn_{0.09}$ double quantum well heterostructure

After validating our measurement approach, we can study samples with heterostructures promising for hosting hole spin qubits. Specifically, we study a sample with a DQW whose heterostructure illustrated in FIG. *3* and growth procedure is described in Ref. [40].

First, at a temperature of 20 K and in the absence of an external magnetic field, we perform laser power-dependent PL measurements to optimize the SNR for our next measurements while ensuring that laser heating does not broaden the PL spectra obtained from the sample.

FIG. *4*a presents the normalized spectra and their fitted Lorentzian functions at different laser powers in the range of 40-130 mW measured before the excitation fiber (orange line in FIG. *1*), which corresponds to 22-73 mW at the sample.

The integrated PL intensity, extracted peak energy, and FWHM are illustrated in FIG. *4*b, FIG. *4*c, and inset of FIG. *4*c, respectively. The integrated PL intensity (FIG. *4*b) demonstrates a power law behavior (dashed lines) with the exponent of $0.70 \pm 0.15$ consistent with Auger-limited recombination models reported for similar GeSn/SiGeSn QWs [40], noting the limited excitation range here. While the integrated PL intensity does not exhibit saturation, we do not apply laser powers greater than 130 mW to avoid overheating the cryostat. Apart from integrated PL intensity, the lack of significant broadening of the FWHM at increasing laser powers (inset of FIG. *4*c) suggests that excitation-induced effects such as state filling or many-body broadening remain weak within the investigated power range. Finally, the negligible change of the peak energy with the laser power (FIG. *4*c) eliminates the possibility of the band structure renormalization (change in the bandgap resulted from lattice constant variation) due to the excitation laser power.

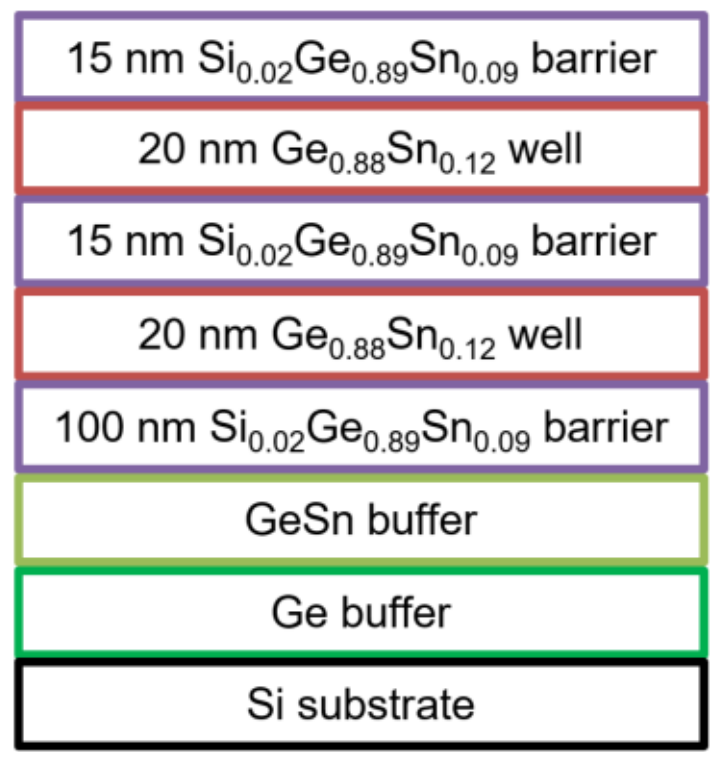

FIG. 3. The schematic of the double quantum well heterostructure.

After ruling out the possibility of laser heating on the spectral behavior of the material under study, we study the temperature-dependent behavior of the DQW with the optimum laser power of 130 mW before fiber coupling (73 mW at the sample), for the minimum effect of heating and maximized SNR.

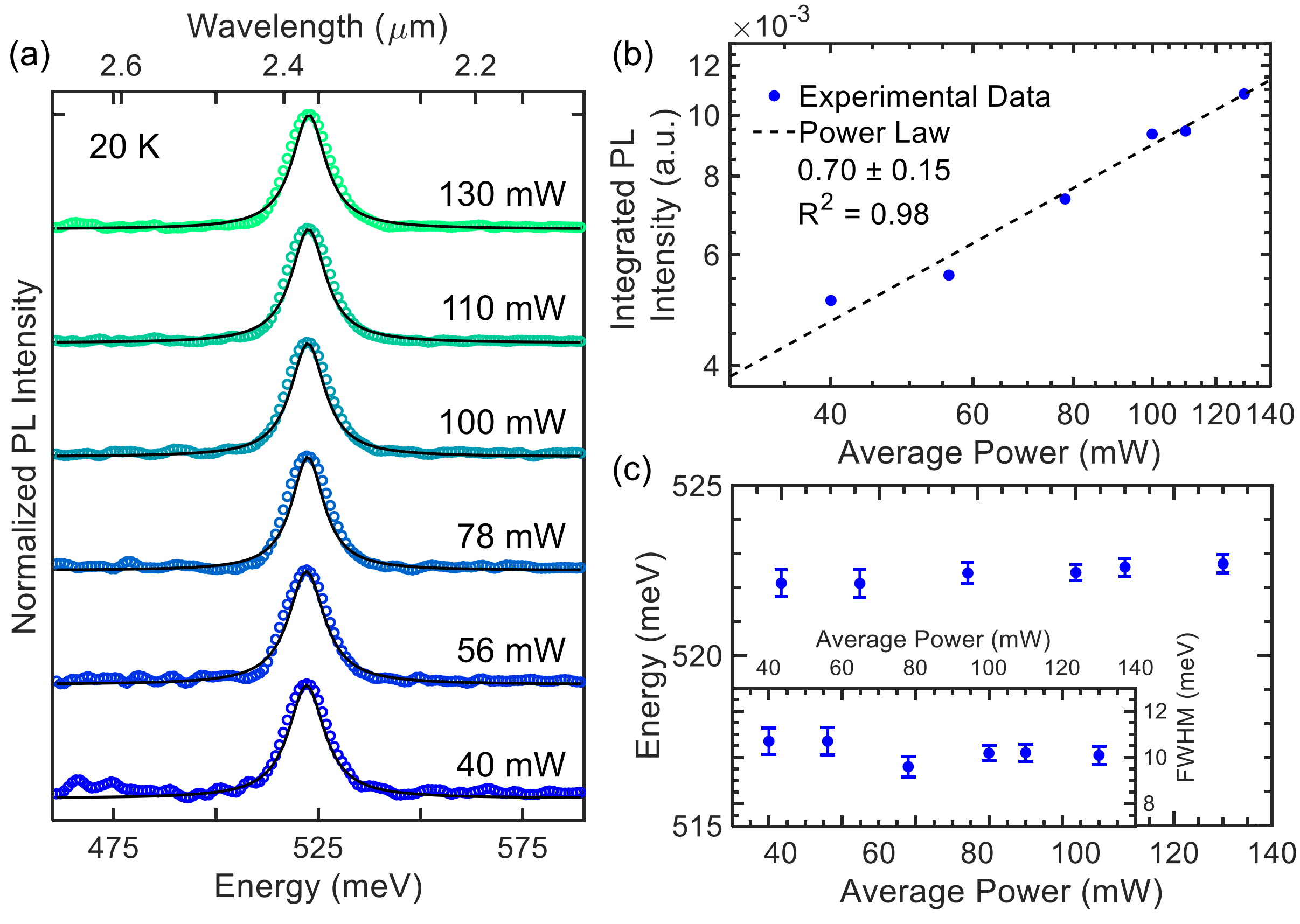

FIG. 4. PL emission from the DQW sample as a function of the excitation laser power at 20 K. (a)The normalized spectra at different excitation powers (circles), fitted to Lorentzian functions (solid lines). (b) The integrated PL intensity as a function of the laser power exhibits no saturation. (c) The photon energy as a function of the laser power. The negligible change of energy implies the lack of band structure renormalization. Inset: The FWHM as a function of the laser power. The unchanged linewidths suggest that the effect of laser broadening is negligible.

We investigate the spectral behavior of the bandgap by performing temperature-dependent PL measurements from the DQW (FIG. *5*). The normalized PL spectra at different temperatures are fit to Lorentzian functions (FIG. *5*a).

As expected, and in a good agreement with previous works [23, 40, 41], the integrated PL intensity decreases as the temperature increases (FIG. *5*b), consistent with the trend of Arrhenius-type (Eq. (4), Ref. [30]) thermal quenching [39]. This trend suggests the onset of thermally activated non-radiative processes, such as carrier escape or exciton dissociation. Deviations from simple power-law behavior indicate that mechanisms beyond Auger-type recombination may be involved, with temperature-enhanced carrier redistribution (e.g., diffusion from the buffer into the quantum wells) also potentially contributing. Furthermore, we observe no change in the FWHM as a function of temperature (inset of FIG. *5*c), indicating no evidence of thermal phonon-related broadening. Meanwhile, a small redshift in the photon energy is observed (FIG. *5*c), amounting to ~ 3 meV over 60 K. This shift is significantly smaller than expected for the thermal-phonon–dominated regime and suggests that the system remains below the Einstein temperature, where the bandgap is only weakly temperature dependent. The modest shift suggests that the temperature-dependent contribution from electron-phonon interactions remains relatively small within this temperature range [42, 43].

### C. Observation of diamagnetic shift and Zeeman splitting at PL spectra from $Ge_{0.88}Sn_{0.12}/Si_{0.02}Ge_{0.89}Sn_{0.09}$ double quantum well heterostructure

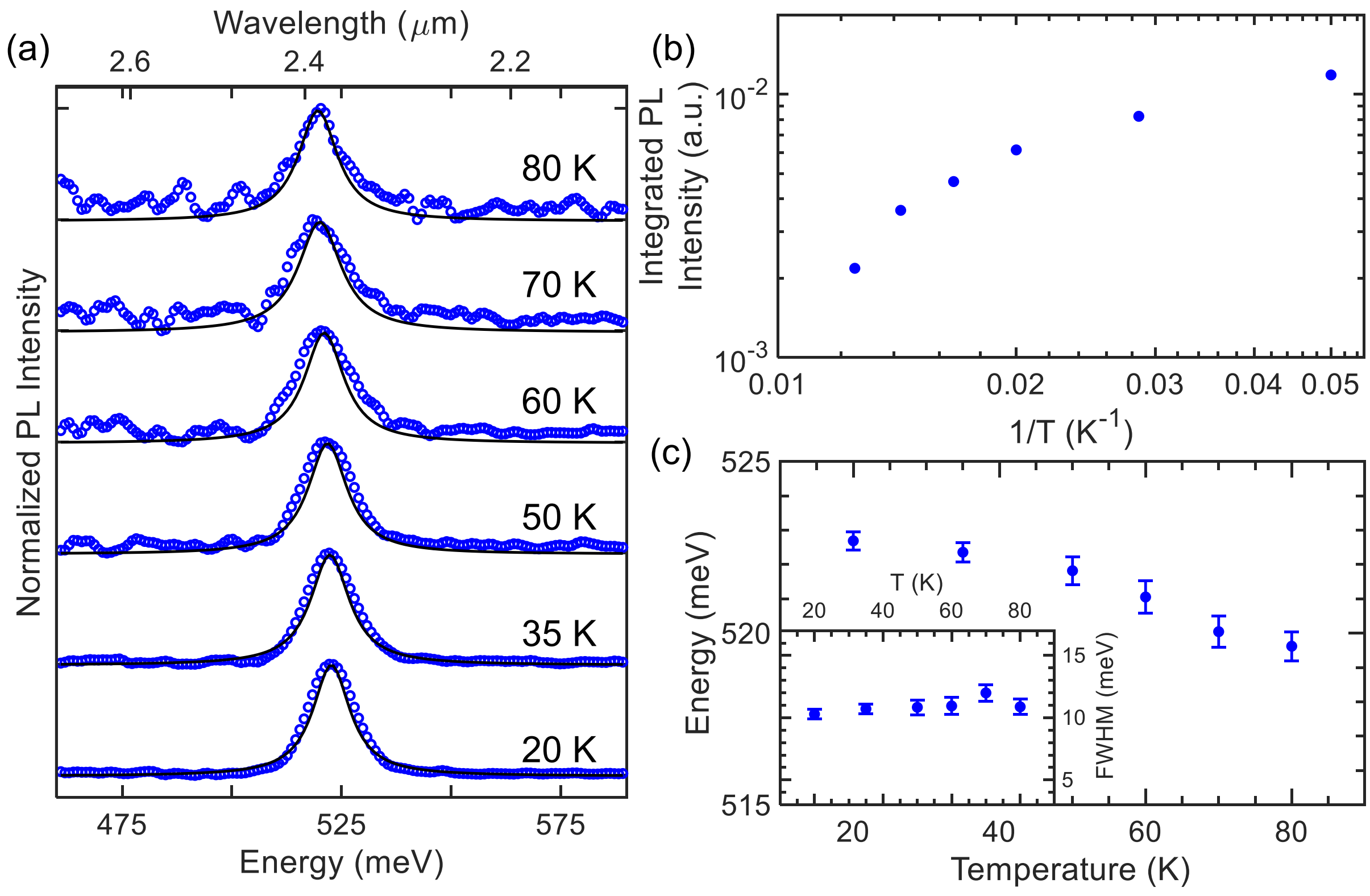

FIG. 5. (a) The normalized temperature-dependent PL spectra from DQW under study (circles) and fit to Lorentzian functions (solid lines) demonstrate an SNR reduction as the temperature increases. (b) The extracted integrated PL intensity as a function of temperature. The decreasing trend with temperature is consistent with the reduction in radiative recombination consistent with direct bandgap materials. (c) Photoluminescence peak energy as a function of temperature exhibiting a minimal redshift (~ 3 meV over 60 K) with increasing temperature. Inset: FWHM as a function of temperature. The constant values indicate that broadening is not dominated by thermal phonons.

To quantitatively investigate the optical properties governing the physical parameters of the DQW [44], we perform magneto-optical PL spectroscopy at a temperature of 20 K for which spectral broadening is not dominated by thermal phonons.

FIG. 6a shows the normalized PL spectra emitted from the DQW at a temperature of 20 K and different external magnetic fields fitted to Lorentzian functions. FIG. 6b illustrates the photon energy as a function of the magnetic field that exhibits separate behaviors under low (0-3 T) and high (3-12 T) external magnetic fields [45]. At low external fields (0-3 T), the extracted photon energy increases parabolically with the field ($B^2$), representing a strong diamagnetic shift of the energy, $E_{Dia}$. By fitting the parabolic trend to the equation $E_{Dia} = E_0 + \alpha B^2$ (solid red line in FIG. 6b), where $E_0 = E_g - E_b$, we extract the difference between the bandgap, $E_g$, and the excitonic binding energy, $E_b$, of $E_0 = 522.76 \pm 0.11\ meV$ as well as the proportionality constant of $\alpha = 0.52 \pm 0.03\ meV/T^2$. Meanwhile, at high magnetic fields, the photon energy increases linearly with the magnetic field, which is associated with the shift, $E_{Lan}$, corresponding to the zero order Landau level, $n = 0$. By fitting the linear trend to the equation $E_{Lan} = E_0 + \beta B$ (dashed black line in FIG. 6b), we extract the difference between the bandgap and the excitonic binding energy of $E_0 = 523.10 \pm 0.55\ meV$ as well as the proportionality constant of $\beta = 1.45 \pm 0.07\ meV/T$. The consistent optical bandgap, the exciton resonant, energy of $E_0 \approx 523\ meV$ extracted in both regimes is in agreement with the extracted values from temperature- and power-dependent measurements (FIG. 4c and 5c). Furthermore, the proportionality constant of the Landau trend is given by $\beta = \hbar e/2\mu$ [46], where $e$ and $\mu$ are the electron charge and the exciton reduced mass, respectively. From the value of $\beta$ extracted from the fit, we obtain the value $\mu = (0.0395 \pm 0.0019)\ m_e$ for the reduced mass of exciton, where $m_e$ is the free electron mass.

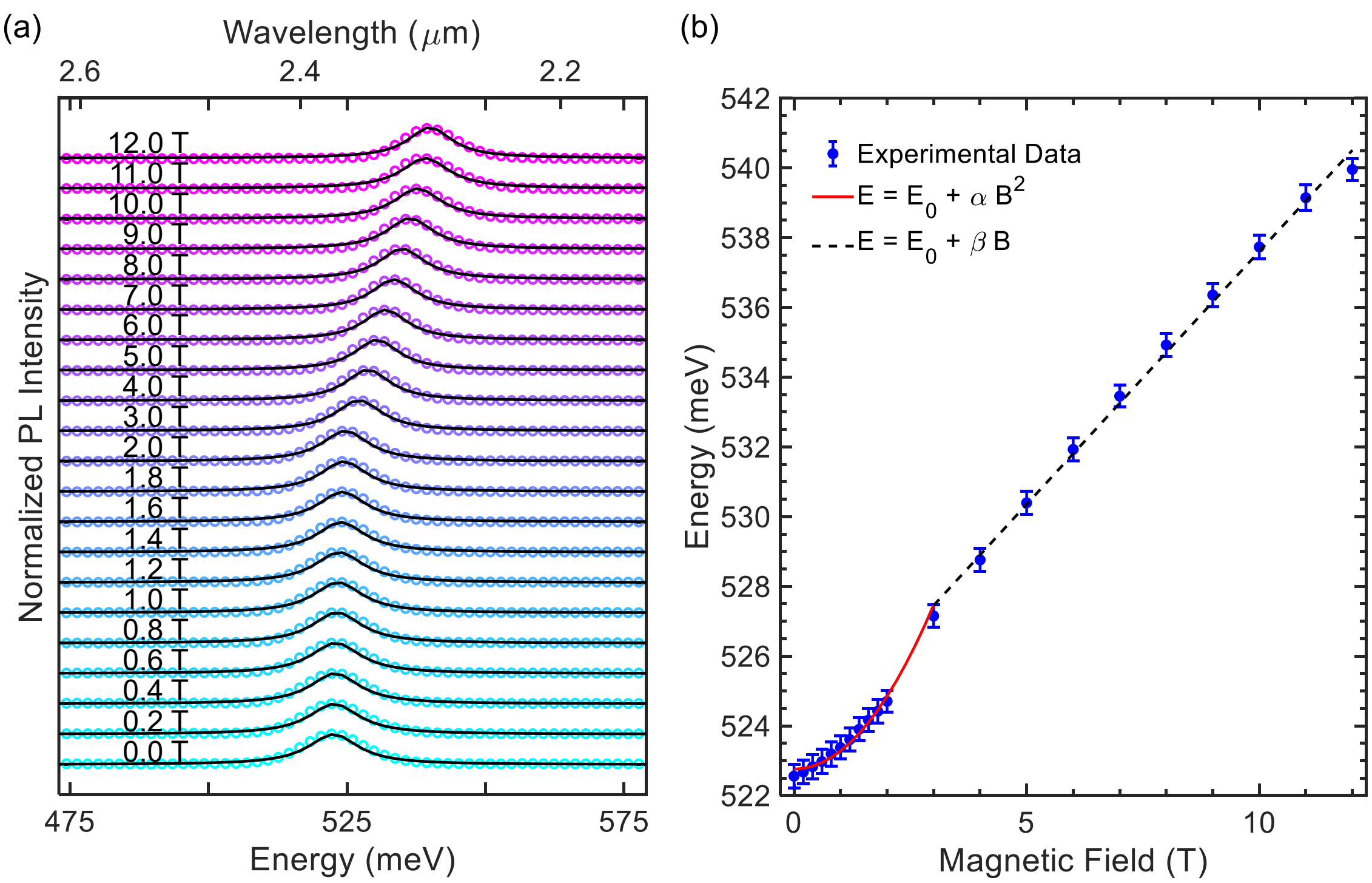

FIG. 6. (a) The normalized PL spectra emitted from the DQW sample as a function of the external magnetic field and their fitted Lorentzian mode; (b) The extracted photon energy as a function of the external magnetic field demonstrates the diamagnetic and Landau shift for low and high magnetic field, respectively.

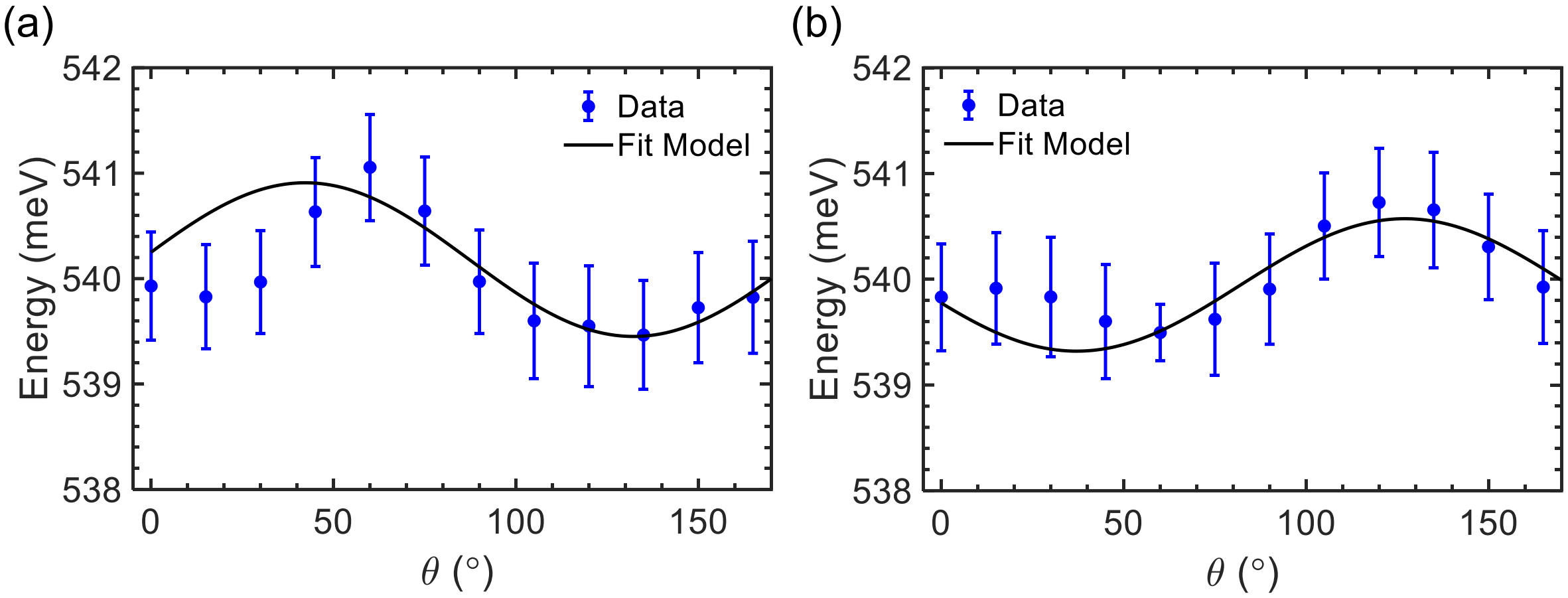

FIG. 7. Polarization-angle resolved PL spectra from the DQW sample at external magnetic fields of (a) 12 T and (b) – 12 T as a function of the angle of the quarter waveplate. The photon energy oscillations at 12 T and – 12 T exhibit chirality as the result of unequal weights of right- or left-circularly polarized components.

Finally, the proportionality constant of the diamagnetic shift is given by $\alpha = e^2\langle r^2\rangle/8\mu$ [47], where $r$ is the exciton in-plane relative position coordinate. Using the calculated value for the reduced mass of exciton and the $\alpha$ extracted from the fit to the diamagnetic term, we can estimate the spatial extent of exciton to be $r_{exc} = \sqrt{\langle r^2\rangle} = 30.7 \pm 1.2\,nm$. We then can estimate the quasi-two-dimensional Bohr radius, considering the thickness of quantum well $20\,nm$, from $a_{2D} = \sqrt{2/3}\,r_{exc}$ [47] to be $a_{2D} = 25.1 \pm 1.0\,nm$, consistent with the large Wannier excitons expected in Ge-based semiconductors. Using the relation $a_{2D} = a_0 \frac{\varepsilon_r}{\mu/m_e}$ with $a_0 = 0.0529\,nm$, we extract a dielectric constant for the GeSn layer of $\varepsilon_r = 19 \pm 1.2$. From this value, the effective Rydberg energy is calculated as $R^* = 13.6\,eV\,\frac{\mu/m_e}{\varepsilon_r^2} = 1.48 \pm 0.19\,meV$. Finally, the corresponding two-dimensional excitonic binding energy is $E_b = 4R^* = 5.9 \pm 0.8\,meV$ [48]. These values, despite the fact that they are model dependent, suggest strong optical transitions, relatively mobile carriers, and potentially long exciton diffusion lengths, and are in good agreement with the reported and expected values for the Ge-based heterostructures. The extracted excitonic binding energy is consistent with previous reports, including values of $5\,meV$ for the shallow bound excitons in SiGe/Si quantum wells [49] and $5.63\,meV$ for excitons in GeSn/Ge multi-quantum-wells [50]. The obtained reduced mass is also comparable to literature values, such as $\sim 0.0427 m_e$ for heavy-hole excitons in the Γ-valley of Ge [51] and larger than that of light-hole excitons in undoped Ge/SiGe quantum wells [52]. Finally, the extracted dielectric constant aligns with theoretical models reported for GeSn quantum dots [53, 54].

In addition to the observed diamagnetic and Landau shifts, the energy of the emitted photons should be affected by the exciton Zeeman shift between two circularly polarized Zeeman-split components $E_\pm = E_0 \pm \Delta E_Z/2$ corresponding to $\sigma^+$ and $\sigma^-$ transitions [55], which we now study using polarization-resolved PL under external magnetic fields of $\pm 12$ T. In our setup, the PL signal emitted from the sample passes through a quarter wave plate and a linear polarizer (FIG. *1*). The orientation of the quarter waveplate with respect to the linear polarizer determines the polarization state of the emitted PL to be analyzed by the FTIR spectrometer. Thus, rotating the orientation of the quarter wave plate allows us to probe the Zeeman components of the emitted PL.

The oscillatory behavior of the polarization-dependent photon energies arises from the standard polarimetric relation between detected intensities and the rotation angle of the quarter-wave plate with respect to the linear polarizer, given by $I_\pm \propto 1 \pm \sin(2\theta + \varphi)$ [56]. When the Zeeman splitting is not spectrally resolved, the measured photon energy corresponds to the intensity-weighted centroid, which can be expressed as $(\theta) = E_0 + (\Delta E_Z/2)\sin(2\theta + \varphi)$.

Motivated by this relation, the polarization-dependent photon energies extracted from the measurements are expected to follow a sinusoidal dependence on the rotation angle. Accordingly, the data shown in Figure 7a and Figure 7b, obtained at external magnetic fields of 12 T and -12 T, respectively, are fitted to an oscillatory function of the form $E = A\sin(2\theta + \varphi) + E_0$. From these fits, the Zeeman splitting is approximated as $\Delta E_Z = 2A$ and the magnitude of the effective g-factor is determined using $|g^*| = \Delta E_Z/\mu_B B$ [57]. We extract the magnitude of the g-factor of $|g^*| = 2.10 \pm 0.35$ for the fit to the experiment performed under an external field of 12 T (Figure 7a) and $|g^*| = 1.80 \pm 0.37$ for the fit to the experiment performed under an external field of -12 T (Figure 7b). These results indicate an effective g-factor of $|g^*| \sim 2$ for the dominant optical transition, providing a benchmark excitonic Zeeman response for GeSn/SiGeSn DQWs; separating electron/hole contributions [58] and any field-induced mixing will require dedicated follow-up measurements.

## IV. CONCLUSIONS

In this work, we used the double modulation FTIR-based PL spectroscopy to characterize quantum materials in the mid-IR region. We first validated our method on a previously studied bulk GeSn sample and then used it to extract the physical properties of a $Ge_{0.88}Sn_{0.12}/Si_{0.02}Ge_{0.89}Sn_{0.09}$ double-quantum-well heterostructure, including the exciton binding energy, spatial extent, and effective exciton g-factor. The observed temperature-independent FWHM suggests that the spectral linewidth is dominated by temperature-independent broadening mechanisms within the investigated temperature range. While the resulting magnitude of the effective g-factor of $|g^*| \sim 2$ quantifies the excitonic Zeeman response, determining the sign and separating electron and hole contributions will require additional experiments/analysis. Furthermore, the extracted values of the binding energy and spatial extent of the excitons are in a good agreement with the literature.

Our approach for studying magneto-PL can be adopted to further investigate the opto-electronic properties of similar materials with different concentrations of Sn, different heterostructure architectures, and the effects of post-growth processing such as hydrogenation or annealing as a procedure to reduce surface-defect-induced disorder and improve the uniformity of the photon emission. By combining PL spectroscopy with optimized material growth and surface treatment procedures to reduce surface-defect-induced disorder and improve the uniformity of optical emission, further enhancement of the PL intensity and hole $g$-factor may enable new material platforms for scalable spin qubits in gate-defined quantum dots and spin-photon interfaces for quantum networking.

## ACKNOWLEDGMENT

We acknowledge funding received from Air Force Office of Scientific Research (Grant Number FA9550-23-1-0302). This work was partially supported by the U.S. Air Force Small Business Innovation Research (SBIR) program under Contract No. FA945325PX004, awarded to JP Analytical, LLC, with NC State as a subcontractor. We thank Dr. Jay Narayan for providing the FTIR spectrometer for performing PL measurements and Paul K. Hinahara from Madison Instruments, Inc. for technical services of the FTIR spectrometer. M. A. thanks Myratgeldi Katyrov for his assistance in implementing lock-in amplification technique, and Mahmoud Attala and Lu Luo for their consultancies on their step-scan method.